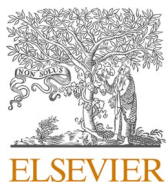

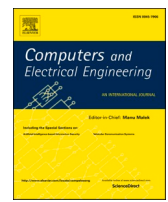

# SAT-CEP-monitor: An air quality monitoring software architecture combining complex event processing with satellite remote sensing☆

Badr-Eddine Boudriki Semlali [a,*], Chaker El Amrani [a], Guadalupe Ortiz [b], Juan Boubeta-Puig [b], Alfonso Garcia-de-Prado [b]

[a] LIST Laboratory, Faculty of Sciences and Techniques of Tangier, Abdelmalek Essaâdi University. The old way of airport, Km 10, Ziaten., PO. Box 416, Tangier, Morocco
[b] UCASE Software Engineering Research Group, School of Engineering, University of Cadiz. Avda de la Universidad de Cádiz, 10, 11519 Puerto Real, Spain

**ARTICLE INFO**

**ABSTRACT**

Air pollution is a major problem today that causes serious damage to human health. Urban areas are the most affected by the degradation of air quality caused by anthropogenic gas emissions. Although there are multiple proposals for air quality monitoring, in most cases, two limitations are imposed: the impossibility of processing data in Near Real-Time (NRT) for remote sensing approaches and the impossibility of reaching areas of limited accessibility or low network coverage for ground data approaches. We propose a software architecture that efficiently combines complex event processing with remote sensing data from various satellite sensors to monitor air quality in NRT, giving support to decision-makers. We illustrate the proposed solution by calculating the air quality levels for several areas of Morocco and Spain, extracting and processing satellite information in NRT. This study also validates the air quality measured by ground stations and satellite sensor data.

## 1. Introduction

Over the last decade, air pollution, climate change monitoring, and natural disaster forecasting have become critical topics in the field of Remote Sensing (RS) monitoring and Big Data (BD) analytics [1,2]. Equally critical and key in the process is the software to be developed for such monitoring and analysis. The great interest in poor air quality is because it can seriously affect human and animal health and negatively impact environmental resources [3]. The World Health Organization (WHO) highlights the influence of air quality in its health reports. The WHO confirms that about 3 million deaths per year are caused by daily exposure to anthropogenic outdoor pollutants, such as Nitrogen Dioxide ($NO_2$), Sulphur Dioxide ($SO_2$), Carbon Monoxide (CO), Ozone ($O_3$) and Particular Matter ($PM_{2.5}$ and $PM_{10}$) derived from the Aerosol Optical Depth (AOD) emitted by industrial, transport and agricultural activities [1]. Thus, monitoring air quality is highly recommended, and this monitoring should be conducted in Near Real-Time (NRT). When high levels of pollution are detected, competent authorities and citizens can be alerted to prevent further damage.

Two main methods are currently used to measure air quality. Firstly, RS techniques can be used from satellite data, which will provide us with large amounts of Trace Gas (TG) data whose density we can evaluate [4] to estimate the air quality. Such approaches have to deal with the complexity of Remote Sensing Big Data (RSBD) due to their different format, enormous volume, various data sources, and high velocity [5]. Secondly, where they exist, we can directly measure the different pollutants' values using the sensors installed in Meteorological Ground Stations (MGS). These values are then used to calculate air quality values. However, there are some limitations; in the case of RS-based proposals, most approaches perform batch RSBD post-processing posterior and do not allow for NRT processing. However, remote data should be treated as fast as possible to guarantee their freshness. In the case of proposals based on MGS, the limitation is evident: air quality cannot be measured in areas without a station. It is worth noting that the deployment of stations is hampered by the difficulty of reaching specific locations or having limited network coverage in those locations and the added cost for the administrations or local governments in charge of deploying them.

In our previous work, some of the authors of this paper successfully proposed SAT-ETL-Integrator [6]: a BD batch processing software to perform the RSBD. The SAT-ETL-Integrator acquires, decompresses, filters, extracts, stores, and integrates refined data into Comma-Separated Value (CSV) files, using the Hadoop Distributed File System (HDFS) [6]. Besides, the other authors of this paper have proposed several software architectures for real-time processing of air pollution values obtained from MGS to warn citizens of health hazards due to poor air quality [7]. Such real-time processing was achieved by integrating Complex Event Processing (CEP) [8] into the proposed software architectures [9,10]. Specifically, this work aims to benefit from both proposals' strengths, combining them in overall software architecture.

In light of the above, we propose the following Research Questions (RQ):

RQ 1: Is it possible to deal with the complexity of RSBD due to their different format, enormous volume, various data sources, and high velocity and then process such data in NRT to guarantee their freshness?

RQ 2: Can we successfully integrate the proposals that have proven useful for estimating air quality through RS techniques with those techniques efficiently used for real-time data processing in MGS?

RQ 3: Can we use the MGS data to validate the data estimated from RS techniques to improve air quality monitoring?

Given our experience in both RSBD analysis techniques and stream processing for real-time decision-making, we firmly believe we can deal with the complexity of RSBD in NRT and successfully integrate proposals effectively used to estimate air quality through RS techniques with those used for real-time data processing in MGS. Additionally, we are convinced that the use of air quality data from the MGS for the validation of data estimated from RS techniques will improve the quality and reliability of air quality monitoring. As a result, this paper provides three main contributions. Firstly, we propose a new software architecture, SAT-CEP-Monitor, which allows us to extract the air pollution data from the satellites using RS techniques and process them in NRT thanks to integrating SAT-ETL-Integrator with CEP. Secondly, we have illustrated the proposed architecture's purpose, estimating the air quality level for several Morocco and Spain regions. The evaluated results show that the new architecture efficiently processed and calculated the air quality in NRT using little storage space. Finally, as the third contribution, we validated the results obtained from SAT-CEP-Monitor in two regions of Spain for one month with the MGS air pollution data available from such regions.

The rest of this article is organized as follows: Section 2 provides background on air quality concepts, platforms, satellite specifications, and sensors. Section 3 describes the process of obtaining information from satellite sensors and discusses CEP. Section 4 describes the proposed SAT-CEP-Monitor software architecture. Section 5 explains how the architecture was used for the present case study and how the CEP patterns were defined. Section 6 provides the results and discusses the experimental analysis. Section 7 then presents the validation conducted between satellite sensor data and the MGS. Subsequently, Section 8 compares our study with related works. In Section 9, we provide answers to the research questions. Finally, Section 10 highlights the conclusions and future research lines.

## 2. Background on Air Quality and Satellite Specification

This section provides background to the impact of air pollution on the environment and details some of the existing air quality monitoring platforms in urban areas (Section 2.1). Secondly, we review the specifications of these satellites relevant to air pollution monitoring and their sensors (Sections 2.2 and 2.3, respectively). Finally, we explain the satellite datasets useful for air pollution monitoring and the unit conversions required for their usage.

### 2.1. Air Pollution and Air Quality Platforms

Air pollution in urban areas increases interest to researchers and political communities [11] due to its negative impact on human health and comfort. Many studies have demonstrated that breathing diseases may occur in peaks of air pollution [12].

Urban air quality has traditionally been monitored with MGS networks equipped with many air pollution and climate sensors. In this sense, many national government projects have been launched to monitor urban air pollution and reduce anthropogenic outdoor pollution's consequences on human health. In the Mendeley dataset linked to this paper (see the Supplementary Materials (SM) section), we can find Fig.s which illustrate Air Quality Indexes (AQI) proposed by Europe, the USA, and Canada (SM File 1).

In order to calculate the air quality level for a particular location, each index requires the most important pollutants to be measured: $PM_{2.5}$, $PM_{10}$, CO, $O_3$, $NO_{2,}$ and $SO_2$. Each air quality level is determined according to a range of values; for instance, the US Environmental Protection Agency (EPA) AQI sets a 51-200 range for moderate air quality. MGS in some countries is scantily distributed. Accordingly, proper mapping of air quality is impossible for such countries [12].

### 2.2. Satellite Specifications

RS techniques require the use of satellites. A satellite is an artificial machine flying in a polar or a geostationary orbit paced at a specific altitude, including the Geostationary Earth Orbit (GEO), Low Earth Orbit (LEO), Medium Earth Orbit (MEO), or High Earth Orbit (HEO). The satellite measures atmospheric, earth, and ocean components using many active and/or passive sensors. Satellite sensors have different Spatial Resolutions (SPR), Temporal Resolutions (TMR), and Spectral Resolutions (STR). After measuring, satellites process values with SPECAN and DOPLER algorithms and transmit data to the ground station to perform data calibration and remove imperfections. The results are data of different processing levels: Level2 (L2) derives the geophysical variables such as the AOD and the Vertical Column Density (VCD) of TG and removes noised datasets; Level3 (L3) maps variables into a uniform geo-temporal grid with greater accuracy.

As the information from the different satellites is enormous, and to avoid details irrelevant to this paper, the rest of the background focuses on satellites that provide environmental information and, in particular, useful to monitor air pollution through RS data [13]. We collected data from the Exploitation of Meteorological Satellites (EUMETSAT) via the Mediterranean Dialogue Earth Observatory (MDEO) ground station installed at the Abdelmalek Essaâdi University of Tangier in Morocco, the Earth Observation System Data and Information System (EOSDIS) of the National Aeronautics and Space Administration (NASA), the National Environmental Satellite, Data, and Information Service (NESDIS) of the National Oceanic and Atmospheric Administration (NOAA), and the Copernicus Open Access Hub (COAH) of the European Space Agency (ESA). All of these are polar satellites passing by a Sun-Synchronous Orbit (SSO). The SSO is a geocentric orbit that combines the altitude and the inclination to cover or pass any point on the planet at the same local time. The satellites named make around 14 scans daily, placed in a LEO with an average altitude of 800 Km.

MetOp data is received from the MDEO station in NRT with a delay of 35 min; TERRA, AQUA, and AURA data are received from the EOSDIS data center of NASA in NRT with an average latency of 80 min and Suomi NPP data are received from the NESDIS of NOAA after one hour of pre-processing in the same unit. In the following paragraphs, we provide further details on the satellites mentioned.

### 2.3. Satellite Sensor Specifications

Table 1 shows the six satellite instruments used for environmental applications. We exploited active sensors illuminating the scanned components with radiation and detecting the object's reflected or backscattered energy. Active sensors' key advantages are that they obtain images day and night and are unaffected by clouds and poor weather conditions. Alternatively, there are passive sensors, which detect sunlight and photons reflected from the scanned object. This kind of sensor provides a complete image of the electromagnetic spectrum. However, they can only work during the day and are affected by adverse weather conditions.

Satellite sensors are also characterized by the SPR, meaning the specific Field of View (IFOV) captured by the instrument during a quick scan or the linear dimension of the ground resolved by each pixel. Thus, all the sensors used have a low SPR, except for the Moderate-Resolution Imaging Spectroradiometer (MODIS) and the Visible Infrared Imaging Radiometer Suite (VIIRS).

The third feature is the TMR, which refers to the amount of time required to revisit and collect data for the same location. All the cited instruments have a high TMR ranging from 12 to 24 hours.

The fourth feature is STR, which is defined as the number of spectrum bands in which sensors can amass the reflected radiance. The sensors used have various Spectral Resolutions (STR), including Infrared (IR), Near- Infrared (NIR), Ultraviolet (UV), and Microwave (MW) [13].

Finally, we were interested in detecting AP, especially in the tropospheric layer. Accordingly, all the selected sensors provide data on atmospheric and tropospheric slices.

### 2.4. Air Pollution and Climate Datasets from Satellites

This section explains various key air pollution variables from satellite sensors where all the satellites are at an altitude of 0 to 2 km. Some satellite sensors measure certain atmospheric variables at a fixed altitude. For instance, the Ozone Monitoring Instrument (OMI) estimates the cloud pressure from where clouds occur. Another example, the MODIS, calculates the Land Surface Temperature (LST) exclusively at ground level. Accordingly, this variable is represented in a 2-Dimensional (2D) axis, where X is the Latitude and Y the Longitude. Our study refers to these variables, represented with 2D and a static altitude, as Standard (S). However, other satellite instruments measure the variables in a Vertical Profile (VP). For instance, the Advanced Microwave Sounding Unit (AMSU) calculates the temperature of the atmosphere in several vertical slices. Thus, this variable could be projected into 3-Dimensional (3D) charts where X is the Latitude, Y is the Longitude, and Z is the altitude in our study. We call this type of variable VP. Alternatively, we can also find a few satellite instruments that quantify the VCD, the Total Column (TC), and the Vertical Mixing Ratio (VMR) of the TG, including $SO_2$, CO, Carbon Dioxide ($CO_2$), Nitrogen Oxides (NOx), $NO_2$, $O_3$, Methane ($CH_4$), and the AOD from which we can derive the value of the Particular Matter ($PM_{10}$ - $PM_{2.5}$). The following paragraphs explain the variables and the required unit conversions.

The VCD and the Slant Column Density (SCD) are the concentration of the TG along the atmospheric light path. Thus, the VCD is the vertically integrated concentration along a path: $\int_s^0 \rho\, ds$. The standard unit of the VCD is Molecules/$cm^2$ or Kg/$m^2$. SM File 2 shows all the useful conversion units for TC and VCD. The VCD of the dry air is $2*10^{25}$ Molecules/$cm^2$, equivalent to 15 800 Kg/$m^2$, and the typical tropospheric VCD of $O_3$ is $8*10^{17}$ Molecules/cm$^2$.

**Table 1**
The most commonly used satellite sensors for air quality and climate monitoring.

| Satellite | Sensor | Type (Active/Passive) | SPR | TMR (Hours) | STR | Layer |
|-----------|--------|-----------------------|-----|-------------|-----|-------|
| AURA | MLS | Sounder (P) | 5 x 500 x 3 km | 24 | MW (118 – 2500 GHz) | Troposphere |
| MetOp | GOME-2 | Spectrometer (A) | 80 x 40 km | 24 | UV (0.26 – 0.51 nm) | Atmosphere |
| MetOp | IASI | Sounder (P) | 12 Km at nadir | 12 | IR (3.6 - 15.5 μm) | Atmosphere |
| NPP | VIIRS | Radiometer (A) | 375 m at nadir | 12 | NIR (0.41 - 12 μm) | Atmosphere |
| TERRA | MODIS | Spectrometer (A) | 250m – 1 km | 24 | IR (0.4 - 14.4 μm) | Atmosphere |
| NPP | OMPS | Hyperspectral (A) | 2.7 km at nadir | 24 | UV (250 – 310 nm) | Atmosphere |

$O_3$ has the Dobson Unit (DU), which is equal to the number of $O_3$ molecules required to make a pure layer of $O_3$ with $10^{-2}$ mm thick at $0\ ^\circ$C of temperature and $1^{atm}$.

The VMR is the TG amount divided by the other mixed constituent's total amount, as shown in Equation 1. Its standard unit is the Mol/mol or the Kg /Kg. This could also be given by the Part per Million of volume (PPMv) unit. Equation 2 explains how to convert the ppm (mass) to PPMv (volume). SM File 3 includes all the Molecular Weights (MW) of the atmospheric gases with their averaged VMR and all the functional conversion units of the VMR.

$$\text{VMR(PPMv)} = \frac{n(Gas)}{n(Total) - n(Gas)} \tag{1}$$

$$\text{PPMv} = \text{PPMm} * \frac{\text{Molecular Weight(Gas)}}{\text{Molecular Weight of dry air } (= 29)} \tag{2}$$

The density of TG in the atmosphere is measured with the VCD and the VMR. The difference between the two is that the VCD is the number of molecules through all the atmospheric layers. However, VMR is the number of molecules in a specific segment of the atmosphere. The VCD can be approximatively converted to the VMR and vice versa. The first step is to calculate the Total Column Averaged (TCA) using Equation 3 and then deduce the VMR using Equation 4.

$$\text{TCA(PPMv)} = \frac{\text{TC or VCD(Kg.m} - 2)}{\text{TC of dry air } (= 15800)} \tag{3}$$

$$\text{VMR(PPMv)} = \text{TCA} * \frac{\text{MW(Gas)}}{\text{MW of dry air } (= 29)} \tag{4}$$

## 3. Background on Technologies for Remote Sensing and Complex Event Processing

This section introduces how satellite data can be used through RS technologies and CEP technology; both are key for SAT-CEP-Monitor architecture implementation.

### 3.1. Using Satellite Data for Remote Sensing

Using satellite sensor data, the location of the high densities of TG ($NO_2$, CO, $SO_2$, $N_2O$, and so on) can be identified with great accuracy. A complete survey of the studied zone can be provided, depicting the significant sources of air pollution (factories, dumps, thermal power plants, road transport, etc.) and indicating the relationships between activities in a city and the distribution of air pollution concentration [14]. Accordingly, RS data permits us to identify where efforts should be made to reduce harmful gas emissions.

Generally, satellite datasets are estimated by several satellite sensors and then transmitted via downlink channels to be received in ground stations, such as the Office of Satellite and Product Operation (OSPO) of NOAA and the EUMETCast service of EUMETSAT. Note that EUMETCast is a multiservice broadcasting system based on multicast technology. It uses geostationary satellites for commercial telecommunications using Digital Video Broadcasting by Satellite - Second Generation (DVB-S2) standards and research networks to transmit files to many users. Each satellite provides datasets for a variable in a specific product file and then transmits through a specific channel. For instance, The MetOp satellite measures the TC of the CO using the IASI sensor and then stores the values inside a product file containing the "trg" flag in its filename. This product file can be downloaded in NRT using the EUMETCast via the "EPS-Africa" channel.

The preliminary processing then produces L1 data by applying calibrations using dedicated software, such as SPECAN and DOPPLER. The Satellite Operation Control Center (SOCC) of NOAA, the Earth Observing System and Operation System (EOS OS) of NASA, and the Central Facilities of EUMETSAT process RS data to provide L2 data, including refined products, notably: atmospheric chemistry, VCD, TC of TG, atmospheric temperature, humidity, etc. [6].

The processed satellite data are stored inside scientific file formats such as the Network Common Data Form (NetCDF), the Hierarchical Data Format (HDF5), and the Binary Universal Form for the Representation (BUFR). BUFR is a binary data format created in 1988 and maintained by the WMO (World Meteorological Organization). Moreover, all meteorological data can be stored, as can their specific spatial/temporal context and other associated metadata. The NetCDF and HDF5 formats share almost the same features providing an interface programming that helps to read and write scientific data, such as temperature and pressure, located on a specific longitude, latitude, altitude, and time.

RS data are stored in sophisticated data centers, such as the EOSDIS of NASA, the NESDIS of NOAA, the EUMETSAT Data Center, and the European Space Astronomy Centre (ESAC) Science Data Centre.

RS data differ according to the satellite resolution, revisit cycle, spectrum, and application area. Thus, RS is undergoing an explosive growth of data. These advances in sensor technologies have also led to an increase in the complexity of RS data, particularly diversity and dimensionality. Moreover, the RS technique deals with a large collection of datasets from multiple sources, high-velocity transmission, veracity, variety, and complexity, making it challenging to employ traditional data processing architecture and models [13]. The challenge includes the acquisition, decompression, storage, filter, extraction, analysis, and visualization of the datasets. What is more, we must consider the multi-source, multi-scale, high-dimensional, and non-linear characteristics of RS data when using RS to analyze and extract information.

## 3.2. Complex Event Processing

CEP is a remarkable technology that allows large amounts of data to be analyzed and correlated in the form of events to detect important or critical situations (complex events) in real-time. CEP supports organizing, aggregating, and correlating huge amounts of heterogeneous data and identifying the cause-effect relationships among the input events to detect serious and critical situations in NRT. An event can be defined as anything that occurs or could occur and as anything that could occur but does not [8]. A situation is an event or a sequence of events that require an immediate reaction; simple events are indivisible and occurs simultaneously; complex events add a higher meaning to the situation described by the simple events. Such events can occur in a distributed environment but can be processed in real-time using CEP.

A CEP engine is needed to receive and analyze input data and provide output to use CEP technology. As previously mentioned, input data are called simple events, which, in our scenario, could be the values of measured pollutants happening at a certain point in time. In order to analyze input events, we need to define the so-called event patterns, which represent logical rules, templates, and conditions that are applied to simple incoming events that allow us to detect when a situation of interest occurs. Such a situation of interest is called a complex event. Therefore, when a pattern is satisfied, a new complex event is created. This way, CEP is performed in three stages: event capture, event analysis, and response. The first consists of the reception of events expected to be analyzed in the CEP engine. The second consists of the process that correlates the incoming events to detect critical or important real-time situations through event patterns. The final one is the notification emitted once a pattern has detected a particular situation.

There are several CEP engines, including, for example, Apache Flink, Siddhi, or Oracle Event Processing. We used the Esper CEP engine; this pioneering open-source engine boasts processing up to 500.000 events per second and provides a specific language to define event patterns. This language is called Event Processing Language (EPL) and has a similar syntax to SQL.

## 4. SAT-CEP-Monitor

This section explains the proposed SAT-CEP-Monitor architecture, describing each stage in the process.

### 4.1. SAT-CEP-Monitor Architecture in a nutshell

**SAT-CEP-Monitor Architecture in a nutshell**
Fig. 1 shows the SAT-CEP-Monitor architecture proposed in this paper. We can see the different stages from RS data acquisition to final notification in such a Fig., as briefly explained in the paragraphs below and then detailed in the following sub-sections.

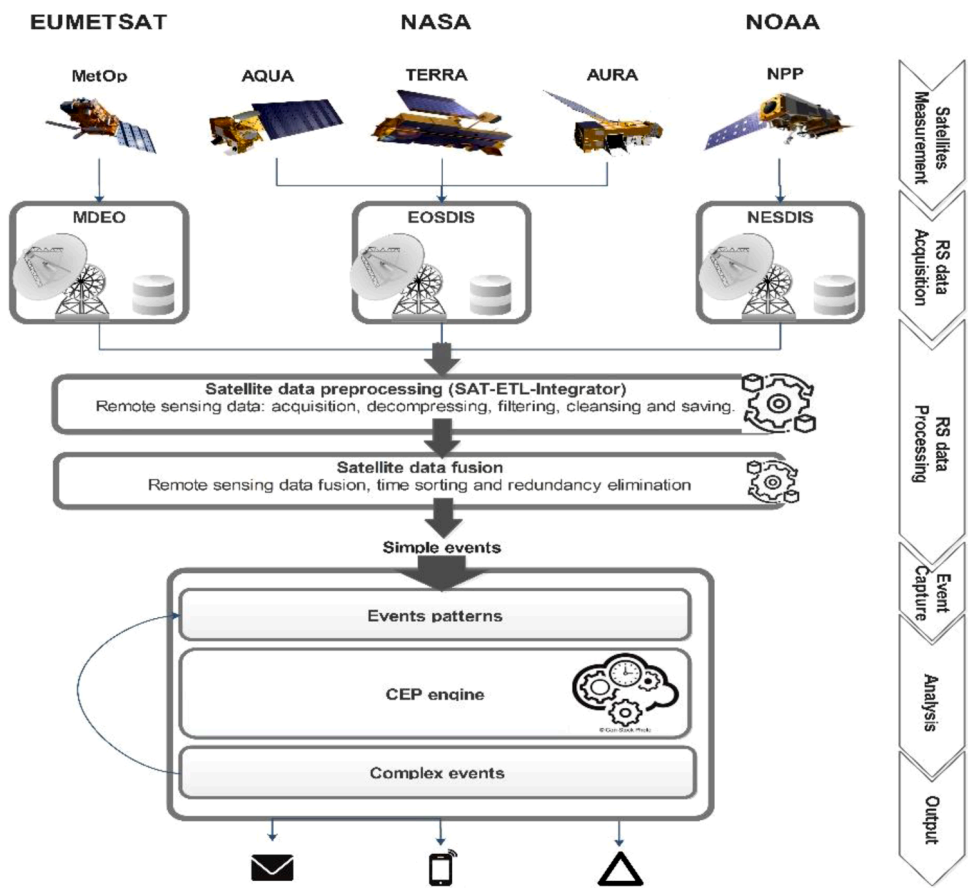

**Fig. 1.** SAT-CEP-Monitor architecture.

**Stage 1. Satellite Measurements:** the data from five satellites —MetOp A&B, AQUA, AURA, TERRA, and Suomi NPP— and six satellite sensors —MLS, GOME-2, IASI, VIIRS, MODIS, and OMPS— are the sources for our architecture. This architecture selects the relevant measurements for air quality monitoring from these satellites [13].

**Stage 2. RS Data Acquisition:** we collected data from the EUMETSAT via the MDEO ground station installed at the Abdelmalek Essaàdi University of Tangier in Morocco, the EOSDIS of NASA, the NESDIS of NOAA, and the Copernicus platform.

**Stage 3. RS Data Processing:** this stage includes data pre-processing —decompression, filtering, conversion, extraction, and finally, data integration [13]— and data fusion —sorting the final CSV outputs by the EpochTime, which are then combined and refined by removing inaccurate and redundant rows. The Final output is regarded as simple events, which are sent to the following stage.

**Stage 4. Event Capture:** this stage is in charge of capturing the events, converting them into the CEP engine format, and submitting them to the latter.

**Stage 5. CEP Analysis:** The events reaching the CEP engine are analyzed according to the EPL patterns deployed in the engine to detect the relevant complex events.

**Stage 6. Output:** the complex events are then notified to the appropriate consumer to be used as the means for air quality NRT monitoring.

### 4.2. Satellite Measurements and RS Data Acquisition

In these two initial stages, we collected data from the European Organization for EUMETSAT, the EOSDIS of the NASA, the NESDIS of the NOAA, and the COAH operated by the ESA.

EUMETSAT is an intergovernmental operational satellite agency with 30 European Member States via the MDEO ground station installed at the Abdelmalek Essaàdi University of Tangier in Morocco. A satellite scans a specified country zone with a particular sensor and then sends the measured values through a downlink to the ground stations via channels. These channels contain many products. Every product includes several variables, such as temperature, pressure, and the VCD of TG. The variables concern 12 levels of altitude ranging from (zero meters) to the middle troposphere of elevation.

The collected data are stored in different file formats, including HDF5, NetCDF, and BUFR of meteorological data. The received data velocity is rapid at 30 000 files/day and has a considerable volume, reaching 60 Gb/day. As a result, their processing is highly challenging in terms of execution time and efficiency [5].

### 4.3. RS Data Processing

The data acquired are first stored and then pre-processed. The pre-processing step includes decompression, filtering of data crossing Morocco and Spain, conversion, filtering based on data quality, filtering that delimits the measured value in a logical range — for instance, the minimum and the maximum value of the pressure must be between 0 atm to 1120 hPa — and, finally, extraction of the refined information and datasets.

We developed the SAT-ETL-Integrator as an Extract-Transform and Load (ETL) software for RS data ingestion and integration. All the details of the SAT-ETL-Integrator can be found in [6]. To illustrate the amount of data that might be obtained by such a procedure, Fig. 2 shows the total number of plots (a single measurement of a variable in a specific time and location in the map) in Morocco and Spain processed in 24h. After the sub-setting, we notice that the number of plots decreases exponentially to keep only values covering the countries' zone of interest. Then, we ensure the data quality, using the minimum and maximum filters to eliminate about 20 percent of inaccurate and erroneous datasets. The final daily number of plots in Morocco and Spain is 10 million and 3 million, respectively.

The pre-processing stage's main outputs are a large number of CSV files, which contain datasets with several variables for each satellite, channel, and product. A final CSV file is composed of 19 columns, which are EpochTime (Unix timestamp format with the instant of measurement), DateTime (date and time of measurement), Longitude (geographic Longitude of the location), latitude (geographic Longitude of the location), and 12 atmospheric levels, as illustrated in SM Files 4 to 8.

A python script was developed to fuse all the final pre-processed CSV files, storing various satellite sensor datasets in a single unified

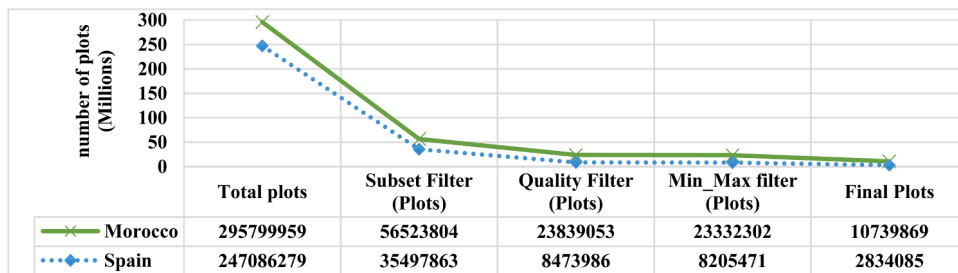

| | Total plots | Subset Filter (Plots) | Quality Filter (Plots) | Min_Max filter (Plots) | Final Plots |
|---|---|---|---|---|---|
| Morocco | 295799959 | 56523804 | 23839053 | 23332302 | 10739869 |
| Spain | 247086279 | 35497863 | 8473986 | 8205471 | 2834085 |

**Fig. 2.** The number of plots after SAT-ETL-Integrator pre-processing.

and combined file containing a specific variable.

### 4.4. Event Capture

To monitor air quality in NRT with the incoming pre-processed and fused RS data, we first read the inputs as simple events and then filter and format the datasets, as shown at the top of Fig. 3. Thus, the first step is to take each row of data in the CSV datasets as single incoming events, including the VCD of CO, $SO_2$, $NO_2$, and $O_3$, and the VMR of CO, $NO_2$, $CH_4$, and $O_3$ and the TC of the AOD. Thus, the VCD variables are converted to the VMR using Equations 3 and 4 in Section 2.4 to have the PPMv unit that fits the EPA AQI standard unit. The second step consists of filtering the empty datasets because we can sometimes find a row containing empty columns representing the 12 levels. Finally, we convert all the integer or float data to double to fulfill the CEP engine proprieties.

Fig. 4 shows that the number of plots increases depending on the SPR of the satellite sensors. The daily total number of simple events in Morocco and Spain is around 1.2 million and 65 thousand plots, respectively.

### 4.5. CEP Analysis

A software module integrating Esper CEP engine was developed for the architecture. This module communicates with a MySQL database, which contains all the required information in seven tables. These tables store all the useful information about the satellites and sensors used, the countries scanned, and the channels, products, variables, qualities, and errors. We also defined and deployed three-event patterns in the CEP engine to detect gas aggregation, pollutant level, and air quality levels. Further details on the particular patterns are given in the case study.

Then, the three services that execute the EPL code are run. The first event pattern consumes the data provided as simple events. Its output complex events, detected by the pattern, are redirected to the second event pattern. The second pattern's output complex events are redirected as input data for the third event pattern. This is represented in the middle of Fig. 3.

### 4.6. CEP Output

The AQ level event pattern serializes the output to a final CSV file, which can be visualized in interactive maps in NRT. All the complex events are plotted into interactive maps to support decision-makers.

## 5. Case Study

This section describes the particular case study used for the experiments, evaluation, and validation of the proposed architecture. First, we describe the steps to be followed to use the architecture in a case study. Second, we establish a case study for a specific

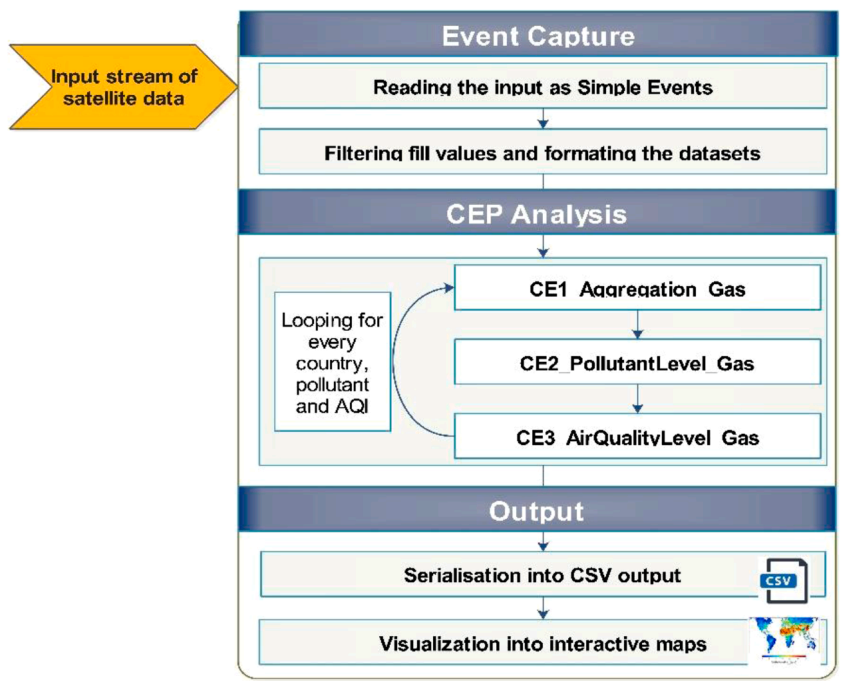

**Fig. 3.** Detailed functionality for CEP stages (stages 4 to 6).

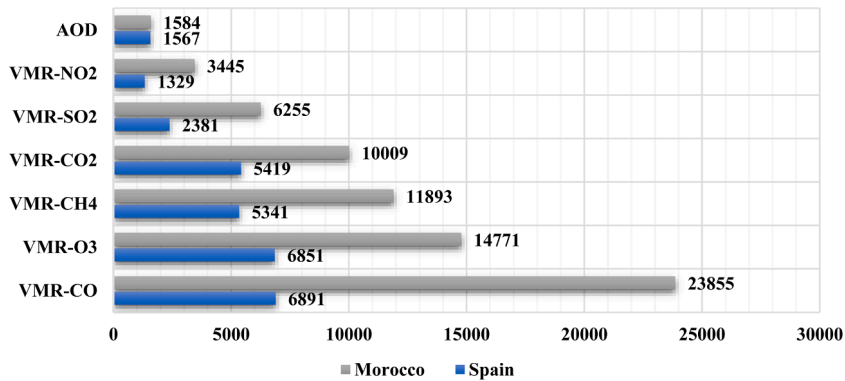

**Fig. 4.** The SAT-CEP-Monitor input data.

location, and, finally, we explain the event patterns implemented.

### 5.1. Using SAT-CEP-Monitor Architecture in a Case Study

We used Linux as the operating system (Debian 9). We also installed Python and Java with additional libraries such as Pandas, h5py, Pyhdf, NumPy, and Esper CEP engine. Subsequently, we deployed the CEP patterns in the CEP engine. The algorithm for RS data pre-processing is explained in [6] and defines the CEP patterns in [15].

The process to use SAT-CEP-Monitor Architecture in a case study can be formalized as follows. Let S be the system's set of satellites of interest $\{s_1,\ldots,s_n\}$, D = $\{d_1,\ldots,d_n\}$ be the set of data of interest from the satellites in S and P be the system's set of pollutant type involved in the study $\{p_1,\ldots,p_n\}$. Also, let D' = $\{d'_1,\ldots,d'_n\}$ be the subset of data of the pollutants of interest and e be the filtered and processed data in D' converted into the format of the event supported by the CEP engine and let C be the system's set of complex event types which might be triggered by the CEP engine $\{c_1,\ldots,c_n\}$. Moreover, the following two different threads $T_1$ and $T_2$ are addressed:

```
Using T₁
for each sᵢ in S
Get data dᵢ from sᵢ
Select P into d'ᵢ from dᵢ
Convert d'ᵢ into e
Send e to the CEP engine
End for each
Using T₂
Wait for any C
Submit cₓ to the output
```

### 5.2. Tailoring SAT-CEP-Monitor Architecture for a Particular Case Study

In the following paragraphs, we explain the particularities for each stage in the architecture.

**Satellite Measurements:** as previously explained, we obtained the data from TeraCast, EOSDIS, and NESDIS.

**RS Data Acquisition:** In particular, we downloaded global data in November 2018. Please note that we chose November 2018 because, for such data, we had full datasets for both satellite sensors and MGS, but any more recent data could have been chosen. We also acquired air quality data from the MGS of the Madrid and Andalusia regions [16] of Spain to validate the air quality calculated from the satellite sensors.

**RS Data Processing:** in this step, the SAT-ETL-Integrator pre-processed only the air pollution data. The total size to process is 23 Gb and 2500 files per day. This batch processing takes around 10 min in a single computer (see the computer features in Section 6.1). This framework also pre-processed 20 Mb of the MGS data in one minute.

**Event Capture:** this stage captures all the events representing Spain and Morocco's air pollution variables for November 2018. It is in charge of capturing the events, converting them into the CEP engine format, and submitting them to the latter.

**CEP Analysis:** we defined and deployed the patterns described in the following sub-sections in the CEP engine. We used the MEdit4CEP [15] graphical model-driven tool to define the CEP patterns, bringing most CEP concepts and features closer to any end-user. Primarily, MEdit4CEP allows domain experts to design event patterns in a user-friendly way. These designed pattern models are automatically transformed into their implementation code, which is executable on a specific CEP engine. In particular, these patterns were deployed in the Esper CEP engine.

**Output:** the complex events are then notified to the appropriate consumer to be used for air quality NRT monitoring.

### 5.3. Event Patterns Defined for the Case Study

As explained in Section 2.1, an AQI, such as the US EPA AQI, measures the air quality level. However, each AQI has its standard ranges for AQ levels. In this case, study, even though we were using air pollution data from Spain and Morocco, we adopted the US EPA AQI because it provides complete information; US EPA AQI levels range from 1 to 6. The US EPA AQI classifies the pollutant level of many TG. It calculates the concerned pollutant's mean concentration across a period (1h, 8h, or 24h) [7]. It can then report the level of pollutants once the average is calculated. The EPA AQI contains six pollution levels ranging from Good to Hazardous; each level has its range. In this study, we processed RS data on seven TGs, specifically: $CH_4$, CO, $CO_2$, $NO_2$, $O_3$, $SO_2$ in PPMv (parts per million by volume), and the PM in µg/m3.

In the following subsections, we explain the patterns that calculate the average pollutant value in the corresponding period of hours (Gas Aggregation). We explain the one calculating the pollutant level according to the averages (Pollutant Level) and, finally, calculating the air quality level, according to the level of all the pollutants (Air Quality Level).

#### 5.3.1. Event Pattern Gas Aggregation

These event patterns compute in 1h, 8h, or 24h slide time intervals, depending on the nature of the gas (see SM File 9) and the aggregate values (Average, Min, Max, and Count) of all the TG (see SM File 10). Since we have seven TG, $PM_{2.5}$, and $PM_{10}$, we implemented seven aggregation event patterns. Fig. 5 shows the conception of this for CO. It reads and loads the simple events and selects only the columns containing the EpochTime, Latitude, Longitude, and the Value of CO. The value must be different from "-0.0", meaning that it is not empty. We also applied a filter, sorting the dataset by Longitude and Latitude. When the event pattern is met, it generates a new complex event called CE1_Aggregation_CO. From the graphical definition made in MEdit4CEP (Fig. 5), we automatically obtain the Esper EPL code for the designed pattern, directly executed in the CEP engine.

#### 5.3.2. Event Pattern Pollutant Level

This event pattern consumes, as shown in Fig. 6, the output of the CE1_Aggregation patterns. It uses the EpochTime, Latitude, Longitude, and the Average concentration of the pollutant to calculate the pollutant level of each TG, $PM_{2.5}$, and $PM_{10}$ based on the EPA AQI. Since we have seven TGs and six air quality levels per TG, together with six levels for $PM_{2.5}$ and $PM_{10}$, we implemented 54 pollutant level event patterns. Fig. 6 shows the pattern for computing the CO pollutant level in good condition. Five more patterns were created for the remaining CO levels. Only the name of the pattern and the range of values in the comparison change, but they all generate complex events named CE2_PollutantLevel. Analogous event patterns were designed for each pollutant and level. This event pattern detects when and where the CO level is good (See SM File 11) in the urban area. When the pattern is met, a new complex event is generated with a stream containing the EpochTime, Longitude, Latitude Level Name, and the Level Number of the pollutant level.

#### 5.3.3. Event Pattern Air Quality Level

Fig. 7 shows the designed event pattern acquiring data from the CE2_PollutantLevel complex events, specifically the EpochTime, Latitude, Longitude, Level Number, and calculates the highest level of all pollutants every 30 min. As a result, a new CE3_AirQualityLevel is produced. Note that the EPA AQI considers that the air quality level is the highest air pollution level of all the pollutants at that moment in the same location (see SM File 12).

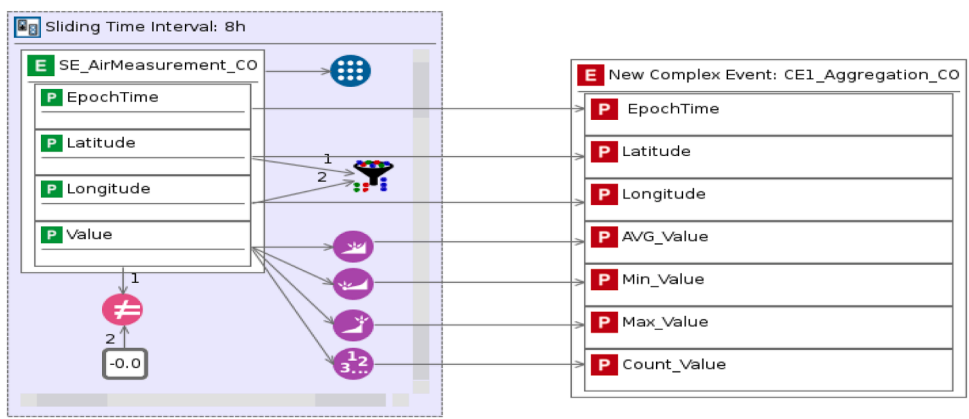

**Fig. 5.** The TG aggregation pattern model.

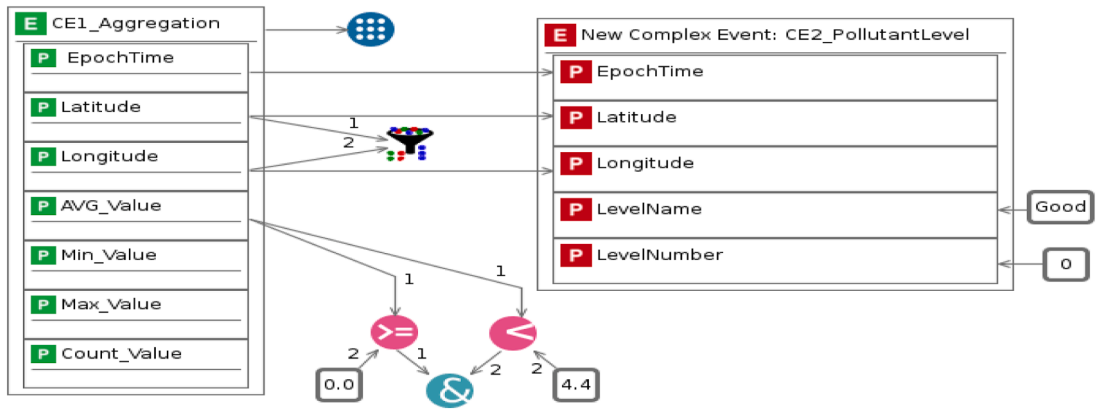

**Fig. 6.** The PL patterns model.

## 6. Experiments and Evaluation

This section describes the experiments conducted according to the description of the case study. First, we provide the benchmarking and metric results obtained in the experiments. We then show the AQI outputs and statistical results, and finally, we visualize and interpret the outputs in interactive maps.

### 6.1. Benchmarking

The experiments were run on a single-node computer with Intel(R) Core (TM) i5-2520U CPU@ 2.50GHz and 16 Gb Random-Access Memory (RAM), running Debian GNU/Linux 10 (64 bit). Table 2 includes the detailed Benchmark of the SAT-CEP-Monitor software architecture during the daily RS data processing. From Table 2, the regular input size in total is 1 Gb containing about 150 000 rows as simple events, including satellite data from Morocco (SAT-Morocco), satellite data from Spain (SAT-Spain), and MGS data from Spain (MGS-Spain). It is worth mentioning that processing in Stages 3 to 6 took less than 17 seconds. Before the CEP processing, the time to download the satellite information is around 25 minutes pre-processing, and fusing it takes less than 10 minutes. This consumes a small part of RAM —around 4 Gb — and uses an acceptable percentage of the CPU —around 60%. The final results are stored in small CSV files below 6 Mb in size, containing approximately 93 000 rows of complex events detected (air quality levels).

### 6.2. AQI Outputs

The final output of every pattern was serialized into a CSV file to be stored, analyzed, and then visualized. SM Files 13, 14, and 15 show the final CSV output of complex events for patterns CO Gas Aggregation, Co Pollutant Level, and Air Quality for a particular day respectively. All these files include the EpochTime, the Latitude, and the Longitude as geo-temporal reference variables of the calculated values, such as the Average value (Gas Aggregation pattern), the Level number (Pollutant Level pattern), and the AQI (Air Quality Level pattern) mandatorily.

Fig. 8 presents the average daily number of AQI plots processed with the SAT-CEP-Monitor software architecture. Note that the number of plots per day in Morocco and Spain is around 35 000 and 13 000, respectively.

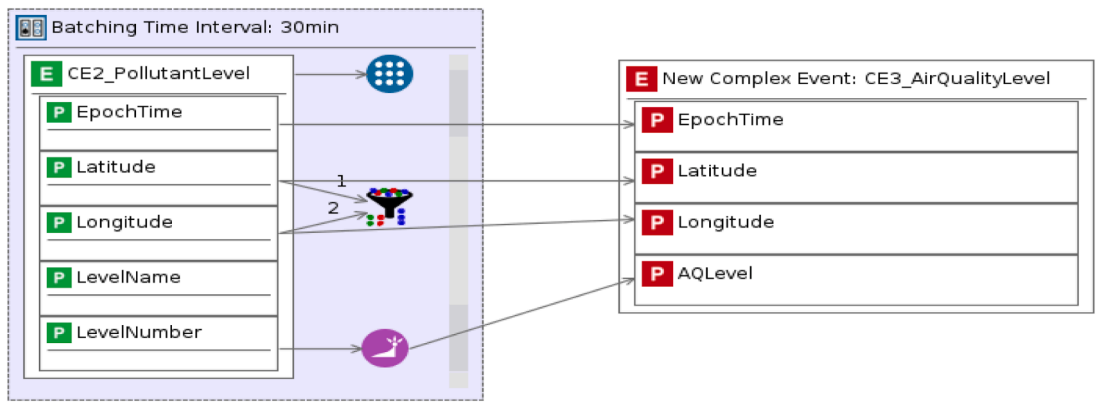

**Fig. 7.** Air quality level pattern model.

**Table 2**
Daily inputs, outputs, and benchmarking statistical results.

|  | SAT-Morocco | SAT-Spain | MGS-Spain |
|---|---|---|---|
| Input size (Mb) | 567 | 322 | 68 |
| Output size (Mb) | 2.9 | 1.5 | 5 |
| Input files | 12 | 12 | 8 |
| Input events | 72 000 | 30 000 | 50 000 |
| Output events | 35 000 | 13 000 | 45 000 |
| Execution Time (s) | 10.5 | 3.1 | 3.3 |
| RAM usage (Mb) | 3 978 | 3 689 | 3 576 |
| CPU usage (%) | 67 | 64 | 62 |

*6.3. Visualizations and interpretations*

The final results of the air quality level are represented in the interactive map of Morocco and Spain. We used the Folium library to manipulate data in Python and then visualize it on the Leaflet map. The folium map enables data to be bound to a map using the heatmap and cycle markers.

Fig. 9 shows the VMR of some TGs with the PPMv and the PPBv units in Morocco on 08/11/2018. The plotted values are generated from the Gas_Aggregation event pattern. We can see that the datasets' range of altitude is between 10m and 200m from the ground. We can see significant CO density in Morocco's northern coastal area (right-hand side of Fig. 9) due to maritime transport activity. Also, there is a high concentration of $CH_4$ around agricultural areas, such as Fes and Larache (left-hand side of Fig. 9).

The left-hand side of Fig. 10 shows that Morocco's air quality is poor in the coastal area due to maritime transport emitting an important density of $NO_2$ and $SO_2$. The same is true for the eastern Spanish coast near Malaga and Valencia (right-hand side of Fig. 10). We also notice that agricultural zones have a high AQI score because they emit $CH_4$ and $CO_2$ due to fermentation. Furthermore, the industrial zones, primarily located in Casablanca and Madrid, emit an extensive amount of CO, $SO_2$, and $O_3$, and thus the AQI is high there. Moreover, urban traffic roads generate a high concentration of $SO_2$, AOD, and PM: Thus, the air quality is poor in large cities, such as Tangier and Barcelona.

## 7. Validation of Satellite air pollution data

SM Files 20 and 21 show the MGS of Andalusia and Madrid, enabling the measurement of the air pollutant concentration in urban areas and some meteorological. The Andalusian regional government network is composed of 61 stations, and the metropolitan network of Madrid includes 22 stations. Each station is equipped with 4 to 8 sensors. The Spanish government provides NRT data to organizations that formally request it every 10 minutes [7]. The MGSs also measure other air pollutants not provided by satellite sensors, such as $PM_{10}$, $PM_{2.5}$, Sulfuric Acid ($SH_2$), Benzene (BCN), Toluene (TOL), and P-xylene (PYX) with the µg/m³ unit. SM File 22 details the average values of the measured VMR of TG, PM, and the AQI in Spain during November 2018.

Validation of data is an operation to compare experimental results from various sources. In this study, we validated the final AQI calculation from satellite sensors and MGS in Spain. Thus, to compare two values, their Global Positioning System (GPS) point must be near. Accordingly, we used Equation 5 to calculate the distance between the two-point GPS in Km. The maximum distance of intersection is 1Km because we worked with satellite sensors with a medium SPR ranging from 1 Km to 5 Km. Moreover, the date and time of the two plots must be equal.

$$D(Km) = Arccos(Sin(LatA)*Sin(LatB) + Cos(LatA)*Cos(LatB)*Cos((LonA) - (LonB)))*R \qquad (5)$$

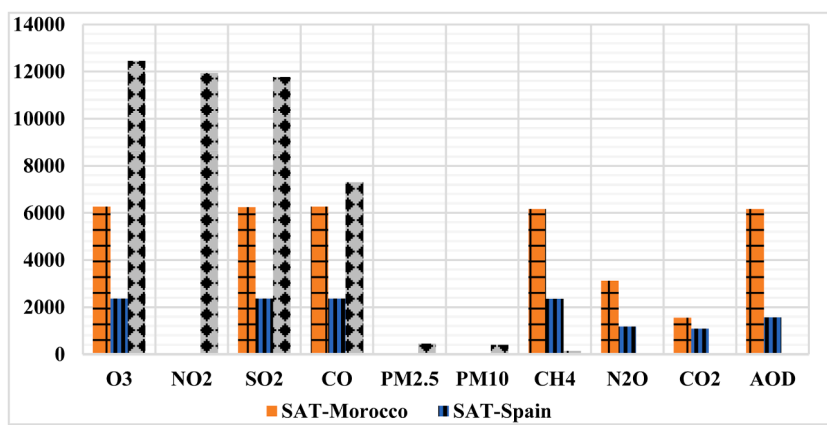

**Fig. 8.** Daily number of the calculated AQI.

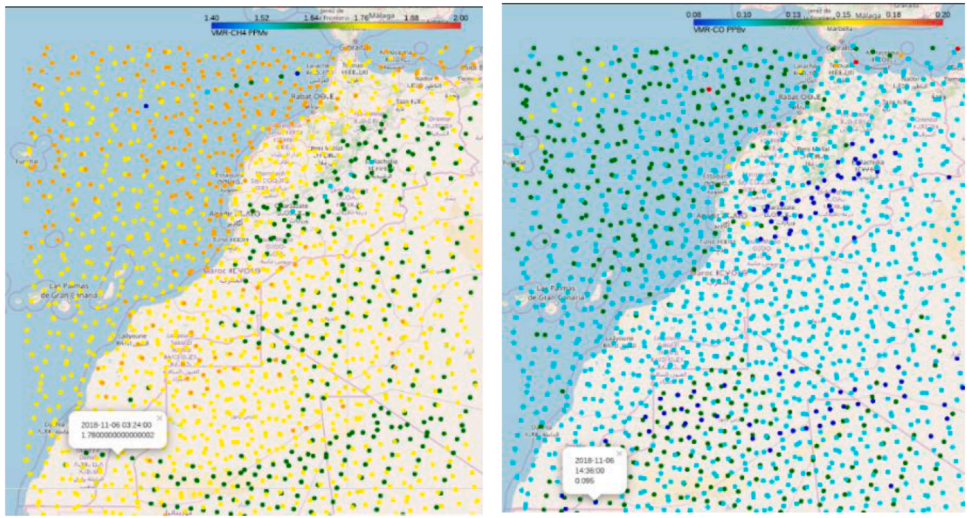

**Fig. 9.** VMR of CH4 and CO in Morocco on 11/08/2018. (see SM Files 16 and 17, respectively)

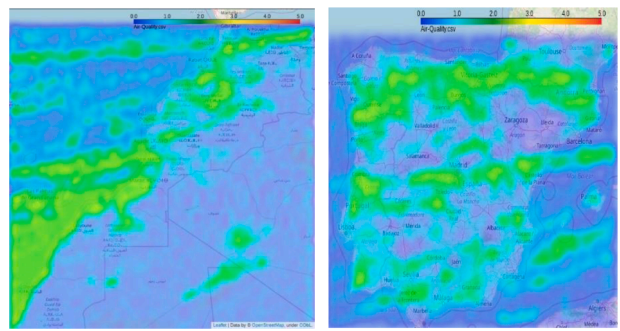

**Fig. 10.** Averaged air quality level in Morocco and Spain from 11/01/2018 to 11/12/2018 (see SM Files 18 and 19)

D is the distance between the two locations in km. LatA and LonA are the GPS coordinate of the first location. LatB and LonB are the second location's GPS coordinates, and R is the Earth's radius, which is equal to 6378 Km.

Fig. 11 shows the validation chart of the estimated AQI from the satellite sensors used and the MGS of Spain during November 2018. We notice that the values of the AQI are close. The coefficient of correlation is 0.75. It is worth noting that this coefficient is acceptable in terms of accuracy and validation. However, it would be enhanced if the distance between the MGS locations and the satellite measurements were closer. For this reason, it is recommended to use other satellite sensor data with a higher SPR. Furthermore, we should apply the techniques of spatial interpolation to obtain additional measurements.

## 8. Related work

In this section, we describe other approaches related to the present methodology. Concerning approaches related to air quality measurements with MGS, we can mention the paper by Rohde et al. [12], where the authors applied the Kriging interpolation for ground station data to calculate the densities of PM, SO₂, NO₂, and O₃ from over 1500. The results showed that the estimated PM2.5 levels from both models largely coincide with those observed, ranging from 85 to 100 µg/m³. Many studies have been conducted to monitor air quality using RS techniques. Of these, we can mention Clerbaux et al. [17], who presented an overview concerning the IASI MetOp satellite sensors and explained how to extract information from the TG atmospheric distribution. Boudriki Semlali et al. [1] also processed MetOp and Meteosat Second Generation (MSG) data. They found a correlation between the emitting sources in Morocco and the VCD of the pollutant. RS techniques can exploit satellite sensors to measure surface earth, ocean, and atmospheric components at a distance through electromagnetic waves [13], but are also useful for environmental applications such as AP, O₃ precursor, and climate-changing monitoring. Currently, the accuracy of satellite sensors is lower than that of ground-based sensors. Thus, combining the two data sources would be the best option to efficiently monitor air quality in urban areas.

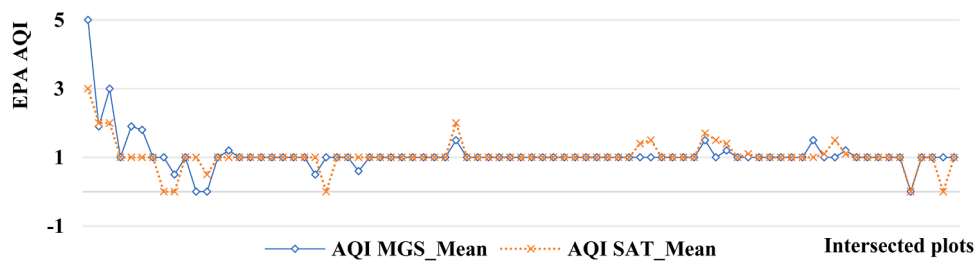

**Fig. 11.** Validation chart of the calculated AQI from satellite sensors and MGS in Spain.

It is also important to mention proposals focused on RSBD: Ma et al. [18] presented a brief overview of the BD and data-intensive problems, including the analysis of RSBD, BD challenge current techniques, and works for processing RSBD. Liu et al. [19] overviewed the characteristics of RSBD and some promising processing techniques and applications. Boudriki Semlali et al. [5] also presented an article showing the main features of the satellite data used in a case study, proving that satellite data are BD. They suggested a Hadoop architecture for pre-processing. These works focus only on batch processing and provide no input concerning streaming processing. As RS should keep its freshness, not providing a stream processing is a weakness for RS data processing.

Some important approaches focused on new architectures for RS processing are explained in the following lines. Xiao et al. [20] developed a novel framework for spatial data processing applied in the National Geographic Conditions Monitoring (NGCM) project in China. Experiments showed that the structure achieves better performance than traditional process methods. Zaytar [11] presented a system to process the BUFR files coming directly from satellite sensors and transform them into a dataset ready for analysis and visualization. Boudriki Semlali [14] developed a Java-based application software to collect, process, and visualize environmental and pollution data acquired from the MDEO platform. All these papers used RS for air quality or climate change applications. However, their primary data sources were very narrow. As a result, the accuracy of their final results was limited.

Some other studies have also processed RSBD in streaming. For instance, Sun et al. [21] described the current situation of RSBD processing and its role in smart cities and discussed streaming processing systems for RSBD. F. G. de Asis et al. [22] proposed a combination of steaming and MapReduce to analyze time-series data from satellite sensors. The results revealed that this method merges the Hadoop and R language and complex process RS time series data efficiently. We can also mention the work of Manogaran et al., who developed a new algorithm to process weather parameters based on MapReduce, including temperature, precipitation, etc., to find a correlation between the weather variation and the incidence of Dengue. The proposed software treated data rapidly and efficiently, helping health decision-makers [23]. These cited architectures process RS data efficiently but require large capacity in the clusters, consuming a large amount of energy.

In parallel to these advanced solutions, there is an emerging technology for stream processing, CEP, which can efficiently process event streams in real-time in any domain. We can also mention the paper from Diaz et al. [24] where an intelligent transport system model based on the CEP is presented. It uses the Andalusian MGS data to decide on traffic regulation and, accordingly, to reduce pollution in urban areas. Since this approach cannot process RSBD directly, it is indispensable to perform RS pre-processing to obtain a refined stream as input.

Finally, we can also mention approaches that focus on spatial-temporal variability measurements for air quality. Mukherjee et al. [25] have compared various measurements from different geolocated sensors for air pollution, particularly the PM. They have also developed a novel technique to measure the PM in urban areas called AirBeam. As a result, the validation demonstrated that the values cross with a high coefficient of correlation. Similarly, in this paper, we monitored AQI, taking into consideration and PM's density. However, we also included satellite data sources to have a consistent input for the model. As results, the validation shows a convergence between the measured AQI from satellite sensors and MGS data,

Table 3 includes a comparison between our proposal and other related works focusing on applying RS techniques for environmental application. This table includes the following features in the comparison: **Input Type** —the type of file where the pollutant data are stored—; **Sources** —the number of satellite or MGS sensors used as input—; **Size** —the total daily size, in Mb, of the data collected from all the used sensors—; **Velocity** —the daily number of files acquired from all the used sensors—; **RS Application** (AP, Climate Change, Other) —the flags concerning the environmental applications of the study—; **RS data processing** (Streaming, Batch, Single, Distributed, Technologies/Tools, Languages) —the flags concerning the type and technologies for data processing—; **Benchmarking** (Execution Time, Used RAM, Used CPU) —the metrics for the execution time and the used RAM for data processing—; **Output Type** —the type of output format —; **Output Size** —the total daily size of output data, in Mb, after the processing— and **GUI** —a flag that indicates whether the application has a Graphical User Interface (GUI) —. Please note that cells with (-) are missing the exact information.

As shown in Table 3, most cited works use CSV or data stream concerning the input type. In contrast, in our study, we also used scientific file formats, notably, the NetCDF, HDF5, and BUFR. On the other hand, most of the presented studies collect data either from satellite sensors or the MGS. In contrast, we acquire data from both sources to have robust input data. As a result, we obtain a combined output result. The streaming processing velocity reaches millions of rows per day. However, the velocity is lower, around thousands of batch processing files. All the cited proposals apply RS in an environmental application, especially for air pollution and climate change monitoring.

Regarding data processing, some studies use batch, and others, stream processing. However, in our study, we use and benefit from the advances in both techniques. Regarding the language of development, we find that Python and Java are the most commonly used in all the studies.

Concerning benchmarking, we also notice that streaming processing software has a short execution time: less than one minute with

**Table 3**

Comparison with related works.

| Feature/Study | SAT-CEP-Monitor | [3] | [6] | [7] | [23] | [24] | [25] |
|---|---|---|---|---|---|---|---|
| Daily RS data input | | | | | | | |
| Input Type | CSV, Stream | BUFR, GRIB | NetCDF, HDF5, BUFR, GRIB, BIN | Stream, CSV, HTML | Stream, Batch, Database | Stream, CSV | Stream |
| Sources | 6 SS 61MGS | 1 SS | 16 SS | 61 MGS | 32 MGS | 100 MGS | 19 MGS |
| Size (Mb) | 600 | 120 | 60.000 | 1 | - | 1 | - |
| Velocity (Rows/day) | 72.000 | 28.000 | 17 million | 50.000 | 50 million | 50.000 | - |
| RS applications | | | | | | | |
| AP | Yes | Yes | Yes | Yes | No | Yes | Yes |
| Climate change | No | No | Yes | No | Yes | No | No |
| Other | No | No | Yes | No | No | No | No |
| RS data processing | | | | | | | |
| Streaming | Yes | No | No | Yes | Yes | Yes | Yes |
| Batch | No | Yes | Yes | No | Yes | No | No |
| Single | Yes | Yes | Yes | Yes | No | Yes | Yes |
| Distributed | No | No | Yes | No | Yes | No | No |
| Technologies/Tools | Esper CEP | - | Hadoop | - | Hadoop | Esper CEP, CPN Tools | - |
| Languages | Java BASH EPL | Python | Java, Python, BASH | Android Firebase | MR, SQL, Spark | EPL | - |
| Benchmarking | | | | | | | |
| Execution time (Min) | 0.5 | 1 | 200 | 0.5 | 1 | - | 1 |
| Used RAM (Gb) | 4 | 4 | 16 | 3 | - | - | - |
| Used CPU (%) | 60 | 100 | 100 | - | - | - | - |
| Daily RS data output | | | | | | | |
| Output Type | CSV, Stream | CSV | CSV | Push alerts | Stream | Stream | Stream |
| Output Size (Mb) | 3 | - | 600 | - | - | - | - |
| GUI | No | No | No | Yes | No | Yes | - |

low RAM and CPU consumption. However, batch processing requires a strong computer configuration.

Concerning the output data, all related works store their output in CSV files or streams. Finally, the GUI is absent in all the proposed solutions, except for the Air4People Android application [7].

Therefore, the originality of the current work compared to other proposals lies in the very fast satellite data processing from various satellites and MGS with an average latency of 40 min. The software also provides accurate AQI output since it combines satellite data with the MGS data processed in NRT with CEP.

## 9. Discussion

We can suggest that our approach enables dealing with the complexity of RSBD and processing such data in NRT. We have seen how, with the proposed architecture —SAT-CEP-Monitor, we can (1) obtain the data from a series of satellite sensors, storing them in the scientific file formats, (2) pre-process them to keep the relevant data, and (3) fuse the data from the different satellites to obtain a CSV for each pollutant to be examined in a geographical area in a short period. Thanks to CEP techniques, we (4) treat all the data as incoming events processed by the engine, according to the defined event patterns, in NRT. In this way, we can make fruitful use of air quality values when valid since they are obtained in NRT.

We have also shown how we integrated the techniques for obtaining RS data from satellite sensors into comprehensive software architecture. These were previously processed in batches, with the techniques for processing complex events in real-time, which, until now, we only used with data obtained directly from the MGS. The integration is not direct, in the sense that a sensor data processing stage is necessary to filter the relevant data while discarding the erroneous data. Thus, we can say that such integration is indeed possible. It is efficient because it takes less than 25 minutes to download 25 Gb of data from the satellites and a maximum of 10 minutes more to pre-process and merge them to be ready for processing in streaming. The latter is fast: the CEP engine processed the data from Spain and Morocco in 17 seconds

We can affirm that the MGS data can be used to validate the data estimated from RS techniques for air quality monitoring. We used the MSG data from the Madrid and Andalusian areas to validate the satellite sensor data obtained for those areas in Spain. As we worked with satellite sensors with an average SPR ranging from 1 km to 5 km, we compared the stations' MGS and satellite sensor values whose location was within a range of 1 km and took the data for the same date time. The results showed that the values were very close for the month under evaluation, with a correlation coefficient of 0.75.

As a result, we can conclude that we have successfully contributed to NRT air quality monitoring from satellite sensor data, which might be of great use for countries or areas without MSG, helping to control air quality and prevent further health damage to the environment.

## 10. Conclusions

Since air pollution has become a significant issue that threatens human health and environmental resources, we propose the SAT-CEP-Monitor in this paper. This software architecture, which integrates batch and streaming processing through CEP technology, permits NRT processing of satellite sensor data. As a result, we can obtain data from the satellites that cover areas of interest and provide valuable and user-friendly information on important pollutants. Additionally, we can obtain and visualize any part of the Earth's air quality data in less than 40 minutes, even if we do not have MGS in such locations.

Once the data is downloaded from the satellite sensors, the proposed solution calculates over 150 000 input rows in 17 seconds and consumes small RAM and storage capacity: 4 Gb of RAM and 5 Mb of the storage space, respectively. Therefore, this software is a novel contribution to help find new solutions to the RSBD streaming processing.

The proposal was illustrated with the air quality data calculated for Spain and Morocco for one month, using recognized air quality references, such as the US EPA. We also validated the AQI from the satellite sensors measurements with MGS data for some Spanish regions.

As future work, we anticipate using additional RS data from other satellite sensors to obtain a significant number of simple events. It would also be interesting to create a GUI, helping inexpert end-users to utilize the solution. A further interesting future work is to develop an application programming interface enabling meteorological forecast agencies to access and share the output of the SAT-CEP-Monitor software.

## Declaration of Competing Interest

None.

## Acknowledgments

This work was partially supported by the Spanish Ministry of Science and Innovation and the European Regional Development Fund (ERDF) under project FAME [RTI2018-093608-B-C33]. The corresponding author thanks the ERASMUS+ KA107 program for the grant and acknowledges the University of Cadiz for the academic supervision and their research facilities, grant number: 2017-1-ES01-KA107-037422 and 2018-1-ES01-KA107-049705. The authors of this work are also thankful to the Andalusian and Madrid regional governments for providing us with the NRT MGS data.

## Supplementary materials

Supplementary material associated with this article can be found in the Mendeley dataset at https://data.mendeley.com/datasets/ddrk9thbcb/1. We have included an index of the SM files in the main folder of the repository.

Badr-Eddine Boudriki Semlali received his master's degree in the computer system and network engineering from the FST Tangier in 2017. He is currently a Ph.D. student working in remote sensing big data and cloud computing. He has authored some peer-reviewed papers, contributed to many international conferences, and benefited from several international scholarships, particularly the Erasmus+, VLIRUOS, and CMN of Murcia.

Chaker El Amrani received his doctoral degree in mathematical modeling and numerical simulation from the University of Liège in 2001. Currently, he is a full professor in UAE, teaching distributed systems. He is the NATO project director, an IEEE senior member, and a Fulbright visiting scholar at George Washington University. His research interests include cloud computing and environmental information systems.

Guadalupe Ortiz obtained her Ph.D. in Computer Science from the University of Extremadura (Spain) in 2007, where she worked from 2001 as an Assistant Professor. In 2009, she joined the University of Cádiz as a tenured Professor in Computer Science and Engineering. Her research interests focus on integrating complex-event processing and context-awareness in service-oriented architectures in the IoT.

Juan Boubeta-Puig is a tenured Associate Professor with the Department of Computer Science and Engineering at the University of Cádiz (UCA), Spain. He received his Ph.D. degree in Computer Science from UCA in 2014. His research interests include real-time big data analytics through complex event processing, event-driven service-oriented architectures, the IoT, and model-driven development.

Alfonso García-de-Prado received his Ph.D. degree in Computer Science from the University of Cádiz, Spain, in 2017, where he has been an assistant professor since 2012. Previously, he had been working as a programmer, analyst, and consultant for international industry partners. His research focuses on context-aware service-oriented architectures and their integration with CEP and the IoT.